\documentclass[11pt]{article}
\usepackage[a4paper,margin=1.0in,footskip=0.5in]{geometry}
\usepackage{authblk}
\usepackage{graphics,graphicx,epsfig}
\usepackage{multirow}
\usepackage{breqn}
\usepackage{subcaption}
\usepackage[round]{natbib}
\usepackage{amssymb,amsmath,amsfonts,eurosym,geometry,ulem,graphicx,caption,color,setspace,sectsty,comment,footmisc,caption,natbib,pdflscape,array,hyperref}

\newcolumntype{L}[1]{>{\raggedright\let\newline\\arraybackslash\hspace{0pt}}m{#1}}
\newcolumntype{C}[1]{>{\centering\let\newline\\arraybackslash\hspace{0pt}}m{#1}}
\newcolumntype{R}[1]{>{\raggedleft\let\newline\\arraybackslash\hspace{0pt}}m{#1}}

\title{The Carbon Premium: Correlation or Causation? Evidence from S\&P 500 Companies}

\author[a]{Namasi G. Sankar}
\author[b]{Suryadeepto Nag}
\author[c]{Siddhartha P. Chakrabarty}
\author[d]{Sankarshan Basu}
\affil[a]{Indian Institute of Science Education and Research Pune, Pashan, Pune - 411008, Maharashtra, India. E-mail: namasivayam.sankar@students.iiserpune.ac.in}
\affil[b]{Indian Institute of Science Education and Research Pune, Pashan, Pune - 411008, Maharashtra, India, and Faculty of Geosciences and Environment, Institute of Geography and Sustainability, University of Lausanne, Geopolis, Mouline, CH-1015 Lausanne, Switzerland. E-mail: suryadeepto.nag@unil.ch/nag.suryadeepto@gmail.com }
\affil[c]{Department of Mathematics, Indian Institute of Technology Guwahati, Guwahati-781039, Assam, India. E-mail: pratim@iitg.ac.in}
\affil[d]{Department of Finance and Accounting, Indian Institute of Management Bangalore, Bengaluru-560076, Karnataka, India. E-mail: sankarshan.basu@iimb.ac.in}
\begin{document}

\maketitle
\begin{abstract}
	
In the context of whether investors are aware of carbon-related risks, it is often hypothesized that there may be a carbon premium in the value of stocks of firms, conferring an abnormal excess value to firms' shares as a form of compensation to investors for their transition risk exposure through the ownership of carbon instensive stocks. However, there is little consensus in the literature regarding the existence of such a premium. Moreover few studies have examined whether the correlation that is often observed is actually causal. The pertinent question is whether more polluting firms give higher returns or do firms with high returns have less incentive to decarbonize? In this study, we investigate whether firms' emissions is causally linked to the presence of a carbon premium in a panel of 141 firms listed in the S\&P500 index using fixed-effects analysis, with propensity score weighting to control for selection bias in which firms increase their emissions. We find that there is a statistically significant positive carbon premium associated with Scope 1 emissions, while there is no significant premium associated with Scope 2 emissions, implying that risks associated with direct emissions by the firm are priced, while bought emissions are not. 

\end{abstract}



\setcounter{page}{0}
\thispagestyle{empty}

\pagebreak \newpage

\doublespacing

\section{Introduction} 
\label{sec:introduction}

A pertinent question in sustainable finance is whether financial markets are pricing carbon risk. Studies often find that firms with higher carbon footprints are associated with excess stock returns that cannot be explained by other relevant factors. Does this imply that investors are demanding higher returns to stay invested in firms that could have large long-term carbon risk exposure? Or is it possible that firms that perform well and offer good returns to investors have less incentive to be environmentally responsible? There are also a host of studies that find conflicting results about the existence of said excess returns \citep{chakrabarty2023risk}. While there are considerable differences in the empirical specifications used in various studies, very few studies make an attempt to establish a robust, causal effect between greenhouse gas (GHG) emissions and stock prices. The primary objective of this paper is to investigate whether the association between firms' carbon footprints and their stock prices is causal, or merely a correlation.  

According to a report published by CDP \citep{cdp2017}, just 50 companies contributed to half the GHG emissions in the world in 2015. Therefore, if we are to not exceed ecological boundaries, firms may need to reduce their energy consumption from fossil fuels in a substantial manner. Such a transition, whether it comes from government regulation such as emissions trading \citep{zhang2010overview}, carbon taxes \citep{rutherford1992welfare, hamilton1994simulating, oswald2023luxury}, carbon rationing \citep{fawcett2004carbon, howell2012living} or voluntary initiatives from the firms, will have negative consequences on the revenue and profit generation, through various channels such as costs of equity capital \citep{kim2015effect}, stranded assets \citep{curtin2019quantifying}, or through indirect effects of reduced consumption \citep{semieniuk2021low}. Entire sectors may exhibit declines \citep{di2013carbon}, and even switching to cleaner production may not achieve positive impacts immediately \citep{cucchiella2017management}. CDP's report also states that of the 224 fossil fuel companies in the sample, 30\% were public-investor owned and 11\% were private-investor owned, implying that low-carbon transitions could imply large risks for investors. Therefore, it is imperative to study whether investors are aware of such risks and whether this manifests as a risk premium or a penalty in stock prices. 

One possibility is that investors are already aware of risks, and have a preference for stocks of companies that are environmentally responsible. \cite{pastor2021sustainable} argue that in equilibrium, these (green) firms would have lower expected returns than polluting (brown) firms, as the stock prices of green firms may be inflated due to investor preferences. However, in the presence of a growing consensus on impending climate change, real markets can hardly be expected to be in equilibrium, and \cite{pastor2022dissecting} find that there exists an abnormal surplus realized return in green firms rather than brown firms, which the authors attribute to an ongoing increase in the demand for green stocks. In contrast, \citet{bolton2021investors} find a positive association between emissions and stock returns and conclude that markets have already priced carbon risk to a certain extent. The authors interpret the association as a premium demanded by investors to continue investing in risky, carbon-intensive firms. They explain that, since carbon-intensive stocks may have significant future risks, an investor would invest in it only if the firm were providing greater short-term gains that may offset the long-term exposure. This implies that if carbon risk were appropriately priced, there would be a ``fair" carbon premium such that the expected abnormal returns would be equal to the expected long-term exposure \citep{nag2023single}. Thus, firms are incentivized to provide an ``abnormal" short-term premium to investors to keep them invested in their shares, which results in a positive carbon premium. 

Despite risks, however, \citet{bolton2021investors} find unexceptional divestments in polluting firms. \citet{krueger2020importance} find that institutional investors, although aware of the financial risks related to climate change, do not think of divestments as an optimal strategy and tend to opt for risk management instead. Risk management, unlike divestments, demands careful analyses and relies more on measures that could adequately quantify the risks. Thus, the existence of a carbon premium would be of particular interest to academics and investors interested in developing strategies for hedging climate risk \citep{andersson2016hedging, roncalli2020measuring, roncalli2021market}.

In general, there is a lack of consensus in the empirical literature with regard to the existence of a positive carbon premium.  In a systematic review, \citet{chakrabarty2023risk} find positive associations between GHG emissions and stock returns in 9 out of 26 studies surveyed, the remaining finding neutral, mixed, or even negative results.\footnote{Nine articles --\citet{oestreich2015carbon,delview,wen2020china, alessi2021greenium,bolton2021global, bolton2021investors, walkshausl2021carbon, gurvich2022carbon}, and \citet{park2022asset} -- estimate positive carbon premia or a negative greenium. Seven articles -- \citet{ravina2019impact, kuang2020carbon, bernardini2021impact, ji2021impact, munoz2021carbon, meyer2022carbon} and \citet{pastor2022dissecting} -- estimate positive greeniums or negative premia. Six articles -- \citet{maddox2018emissions, zhang2018carbon, gorgen2020carbon,brunborg2021impact, aswani2022carbon, vymyatnina2022green} and \citet{zhang2022carbon} -- find no premia. The remaining four studies -- \citet{mazzarano2021carbon, santi2021carbon, witkowski2021does} and \citet{lo2022measuring} report mixed results.} This lack of consensus in the literature could, in part, be explained by different policies in different markets studied. For instance, several studies that survey European markets \citep{oestreich2015carbon, ravina2019impact, bernardini2021impact} would estimate risks in the context of the European Union's Emissions Trading Scheme, where the presence of carbon allowances may induce effects in the market which may not be replicable elsewhere, especially in markets such as those of the United States or other emerging economies that do not have such regulations in place. Strangely, findings do not agree even when the study is set in the same market circumstances. \citet{oestreich2015carbon} finds a statistically significant positive effect associated with the dirty-minus-clean portfolio, while \citet{bernardini2021impact} and \citet{ravina2019impact} find the opposite effect. \citet{chakrabarty2023risk} also find that there may exist methodological biases in determining whether a study reports a positive carbon premium or otherwise, and the necessity of more studies that explore causality in the estimation of carbon premia.

We use publicly available data on GHG emissions and company financials of 141 S\&P 500-listed firms traded in US markets to analyze whether a larger volume of carbon emissions leads to higher investor returns at present. Our final sample of 141 firms together constitute about a third of the total market capitalization of US markets. We study stock price movements between 2018 and 2019 so that inferences are not confounded by market shocks induced by the COVID-19 pandemic in 2020 and 2021. If the value of a stock is thought of as the demand for the stock, the carbon premium can be thought of as the carbon elasticity of demand. Studying the elasticity of demand tells us how stock prices respond to a one percent change in GHG emissions. Studying the demand curves for the shares of a company against the environmental price would require knowledge of a company's potential returns for a variety of GHG emission levels, which cannot be estimated in a market that is not in equilibrium, with public sentiment constantly changing. Accordingly, we study a panel of firms so as to study the demand for shares at a market level. This poses additional challenges related to confounding variables and firm-fixed effects. To address these, we employ a difference-in-differences specification, controlling for a variety of financial parameters that could independently influence stock prices. 

A crucial assumption in difference-in-difference estimation is that of random treatment \citep{roth2023s}, which is unlikely to be the case in firms' GHG emissions. \citet{bolton2021investors} find that a company's carbon emissions are determined by various factors such as size, return on equity, etc. Under such circumstances, whether a company increases its emissions or by how much a company increases its emissions, i.e., the ``treatment"\footnote{Treatment here is a change in GHG emissions, while control is no change. Since changes in firm emissions can be any real number, we essentially have multiple treatment categories or rather continuous treatment doses.} would be affected by various observables, not controlling for which would result in selection bias in the assignment to treatment and control groups \citep{tucker2010selection}. Endogeneity in treatment assignment may make it difficult to make causal inferences as the effect of GHG emissions may also be a manifestation of the effects of the factors that affect a firm's GHG emissions. Very few studies in the literature have controlled for selection bias in the past. \citet{wen2020china} and \citet{santi2021carbon} both use matching techniques to address selection bias. One drawback of matching techniques is that not all treated firms need to have a corresponding control firm with similar pre-treatment characteristics. This leads to a contraction of the sample size. Furthermore, even if pre-treatment covariates are similar, they may vary differently in the study period. Therefore, following \citet{hirano2004propensity}, we weight our firms in the regressions with the inverse of their propensity scores, to ensure that the assignment to treatment, given covariates, is ''as good as random". Weighting by the inverse of propensity scores simply controls for selection bias by balancing the representation of pre-treatment covariates in treatment and control groups, and since a regression follows, differences in changes of other variables can be controlled for. 

To quantify firms' emissions, we use Scope 1 and Scope 2 emissions. Scope 1 emissions are those emissions that are directly produced by the firm on its premises or on assets controlled by the firm. Scope 2 emissions are the emissions from purchased electricity. These emissions are not produced, but rather consumed by the firm and hence add to the footprint. We do not include Scope 3 emissions because these are reported less frequently and include various different types of emissions, the contributions of which to a premium if there is one, would not necessarily be the same. However, one type of Scope 3 emissions, namely, direct use-phase emissions, is an extremely important category. These are emissions that arise from the products sold by the firm. This is particularly relevant for companies that produce, store or transport fossil fuels. In the event of a low-carbon transition, fossil fuel companies' revenues would be hurt not only because of the emissions they consume but also from reduced consumption of the emissions they sell. We incorporate a variable in our model that proxies for sold emissions of fossil fuel companies, and estimates the risks associated with sold emissions. 

From our analysis, we find a statistically significant, positive carbon premium associated with Scope 1 emissions between 2018 and 2019, although we do not find such a premium for Scope 2 emissions. We find that firms with more assets, fewer liabilities, and low existing emissions were most likely to increase their Scope 1 emissions. Our estimates after weighting by propensity scores, in order to remove selection bias, strengthens the estimate of the carbon premium in both size and significance, implying that the carbon premium is larger for firms that are less likely to increase their emissions. We also find that carbon premia are not symmetric, and that while increases in emissions result in increases in an abnormal increase in stock prices, a decrease in emissions does not lead to a decrease in stock prices. This may be because carbon risk is already valued at the market, and the market expects production to become substantially cleaner, and the pace at which firms are reducing their emissions may not be fast enough. However, an increase in emissions may be unexpected among growing concerns on climate change, and leads to an increase in the valuation of transition risk for said firms, causing investors to demand greater returns on the short term. 

The paper is organized as follows. In the following section, we present the data sources and the descriptive statistics of the data under consideration. In Section \ref{sec:method}, we present the methodology, with the results and main conclusions being presented in Section \ref{sec:results} and \ref{sec:conclusion}, respectively. 

\section{Data} 
\label{sec:data}

We use free, publicly available data from S\&P 500-listed firms for our analysis. Based on availability, we collected emissions data for 209 companies listed in US markets, which are a part of the S\&P 500 index. This data was collected from voluntary company disclosures, either to third-party organizations such as CDP, which the companies themselves also published or from ESG reports of the companies. The Greenhouse Gas Protocol classifies carbon emissions into three categories, namely, Scope 1, Scope 2, and Scope 3. Scope 1 emissions include emissions produced by the firm on its own assets, while Scope 2 emissions include emissions from purchased electricity. The latter are not emissions produced by the company, but rather adding to the company's carbon footprint. Scope 3 is a more heterogenous category that includes a variety of different types of emissions such as emissions from the consumption of services such as those from business-related travel to the emissions arising from the products manufactured by companies. We restrict our analysis to Scope 1 and Scope 2 emissions. This is for two major reasons. First, the number of firms that have data available on Scope 1 and Scope 2 emissions greatly outnumber the number of firms that have data on Scope 3 emissions (in addition to Scope 1 and Scope 2 emissions) and if we were to include firms that had data on all three types of emissions, our sample would shrink considerably. Secondly, while Scope 1 emissions clearly include only emissions produced by the firm, and Scope 2 emissions include only emissions consumed by the firm, this is not true for Scope 3 emissions, which could include both, and there is little reason to believe that any risk would be homogeneously linked to different kinds of Scope 3 emissions. This makes the interpretation of any quantitative results difficult, and a proper analysis would require more clarity while looking within Scope 3 emissions. However, this is beyond the scope of this work. Most of the companies in our database had data for Scope 1 and Scope 2 emissions for the years 2018 and 2019, and we selected firms that had data for both years.
\begin{table}[t]
\centering
\resizebox{\textwidth}{!}{\begin{tabular}{cccccccccccc}
\hline 
&&&&&&&&&&&\\
& \multicolumn{11}{c}{Means (standard deviation)}  \\
\\ \cline{2-12}
&&&&&&&&&&&\\
& \multicolumn{3}{c}{2018}&&\multicolumn{3}{c}{2019}&&\multicolumn{3}{c}{$\Delta$}\\
&&&&&&&&&&&\\
\cline{2-4} \cline{6-8} \cline{10-12}
&&&&&&&&&&&\\
GICS   & Log Scope 1 &Log Scope 2 &Log Total && Log Scope 1 &Log Scope 2 &Log Total && Log Scope 1 &Log Scope 2 &Log Total \\
Sector   & Emissions&Emissions&Emissions&&Emissions&Emissions&Emissions&&Emissions&Emissions& Emissions\\
&&&&&&&&&&&\\
\hline 
&&&&&&&&&&&\\
All Sectors &  12.6607     &  12.3149     &  13.7497      &&  12.6715       &  12.1249       & 13.7022       &&  0.0108    & -0.1900
&   -0.0476      \\
&   (3.1822)    &   (2.1451)    &   (2.4059)     &&   (3.1178)      &   (2.3150)      &   (2.4012)     &&   (0.3198)   & (1.0537)
&     (0.3072)    \\
&&&&&&&&&&&\\
Energy &  15.8383  &   13.9467    &  16.0241      &&    15.7922     &  13.6064       &   15.9206     &&   -0.0462   & -0.3403
&   -0.1035      \\
&   (1.5379)    &   (1.4383)    &   (1.4946)     &&   (1.5201)      &   (1.4846)      &   (1.5067)     &&   (0.2428)   & (0.6967)
&     (0.2532)    \\
Materials & 14.6101   &    14.0287   &   15.1028     &&   14.6046      &   13.9630      &   15.0806     &&   -0.0056   & -0.0657
&    -0.0222     \\
&   (1.5744)    &   (1.3040)    &   (1.4564)     &&   (1.5856)      &   (1.3180)      &   (1.4688)     &&   (0.0635)   & (0.0741)
&     (0.0521)    \\
Industrials & 13.5263   &   12.2232     &    14.2680    &&    13.4607     &            12.1402     &  14.1862    && -0.0656
&    -0.0830  &  -0.0818 \\
&   (2.6229)    &   (1.4218)    &   (2.0105)     &&   (2.6653)      &   (1.3666)      &   (2.0525)     &&   (0.1182)   & (0.1350)
&     (0.0987)    \\
Consumer Discretionary &   12.2436 &   13.4256    &    13.7077    &&  12.2387       &  13.3919       &   13.6832     &&   -0.0049   & -0.0338
&   -0.0245      \\
&   (1.3523)    &   (1.1724)    &   (1.2055)     &&   (1.3346)      &   (1.1955)      &   (1.2176)     &&   (0.0933)   & (0.0485)
&     (0.0434)    \\
Consumer Staples &  13.3777  &  13.4810     &   14.2956     &&   13.3615      &       13.0902  &    14.0090    &&  -0.0162    & -0.3907
&   -0.2866      \\
&   (1.8078)    &   (1.9157)    &   (1.9099)     &&   (1.8213)      &   (1.3867)      &   (1.6306)     &&   (0.0371)   & (0.8564)
&     (0.6171)    \\
Health Care& 11.2774   &    11.3483   &   12.3596     &&   11.2949      &  11.3375       &   12.3321     &&  0.0175    & -0.0109
&   -0.0274      \\
&   (1.5621)    &   (2.5807)    &   (1.3661)     &&   (1.4997)      &   (2.5098)      &   (1.3814)     &&   (0.1542)   & (0.2349)
&     (0.1761)    \\
Financials &  8.5815  &   10.2833    &    10.5032    &&  8.5374       &    10.1018     &  10.3614      &&   -0.0441   & -0.1815
&   -0.1418      \\
&   (1.2514)    &   (0.6718)    &   (0.7265)     &&   (1.2627)      &   (0.7571)      &   (0.7653)     &&   (0.1737)   & (0.2104)
&     (0.1343)    \\
Information Technology&  9.9745  &   11.5278    &   12.0444     &&   10.2007      &    11.6759     &     12.1787  &&   0.2262   & 0.1481
&   0.1343      \\
&   (2.9704)    &   (2.2317)    &   (1.8110)     &&   (2.8408)      &   (2.1638)      &   (1.9971)     &&   (0.6995)   & (0.8332)
&     (0.5739)    \\
Communication Services& 13.8350   &   15.7170    &   15.8588     &&    13.8064     &   15.5690      &   15.7274     &&  -0.0286    & -0.1480
&   -0.1314      \\
&   (0.0000)    &   (0.0000)    &   (0.0000)     &&   (0.0000)      &   (0.0000)      &   (0.0000)     &&   (0.0000)   & (0.0000)
&     (0.0000)    \\
Utilities &  15.9515  &   12.3496    &   16.3178     &&  15.8799       &  11.4556       &  16.2275      &&  -0.0717    & -0.8940
&     -0.0903    \\
&   (2.2097)    &   (1.9429)    &   (1.5875)     &&   (2.1826)      &   (3.2777)      &   (1.5731)     &&   (0.1285)   & (2.3904)
&     (0.0922)    \\
Real Estate &  10.2411  &  12.0469     &  12.3568      &&  10.2890       &   11.9954      &    12.3250    &&   0.0479   & -0.0515
&   -0.0318      \\
&   (0.6985)    &   (1.7374)    &   (1.4855)     &&   (0.6354)      &   (1.6382)      &   (1.3712)     &&   (0.0970)   & (0.1048)
&     (0.1198)    \\
&&&&&&&&&&&\\
\hline 
\end{tabular}}
\caption{Sector wise GHG emissions of firms in 2018 and 2019. Includes 141 firms from 11 sectors. Averages are unweighted.}
\label{tab:sector_emissions}
\end{table}

For stock price data, we use data from Yahoo Finance, and for company financials, we use data from Google Finance, which has data on various variables of interest such as revenue, expenses, profits, earnings per share, etc. Both the databases, namely, Yahoo Finance and Google Finance are free and publicly available. Based on the availability of Scope 1 emissions, Scope 2 emissions, and data on a variety of financial variables, we construct a panel of 141 firms that maximizes the number of variables that we can use for our analysis without compromising severely on the size of our sample. To classify firms by sectors, we use the Global Industry Classification Standard developed by MSCI and S\&P Dow Jones indices, which classifies industries into 11 sectors -- Energy, Materials, Industrials, Consumer Discretionary, Consumer Staples, Health Care, Financials, Information Technology, Communication Services, Utilities, and Real Estate.
 
\begin{figure}[t]
\begin{subfigure}{.5\textwidth}
\centering
\includegraphics[width=1\linewidth]{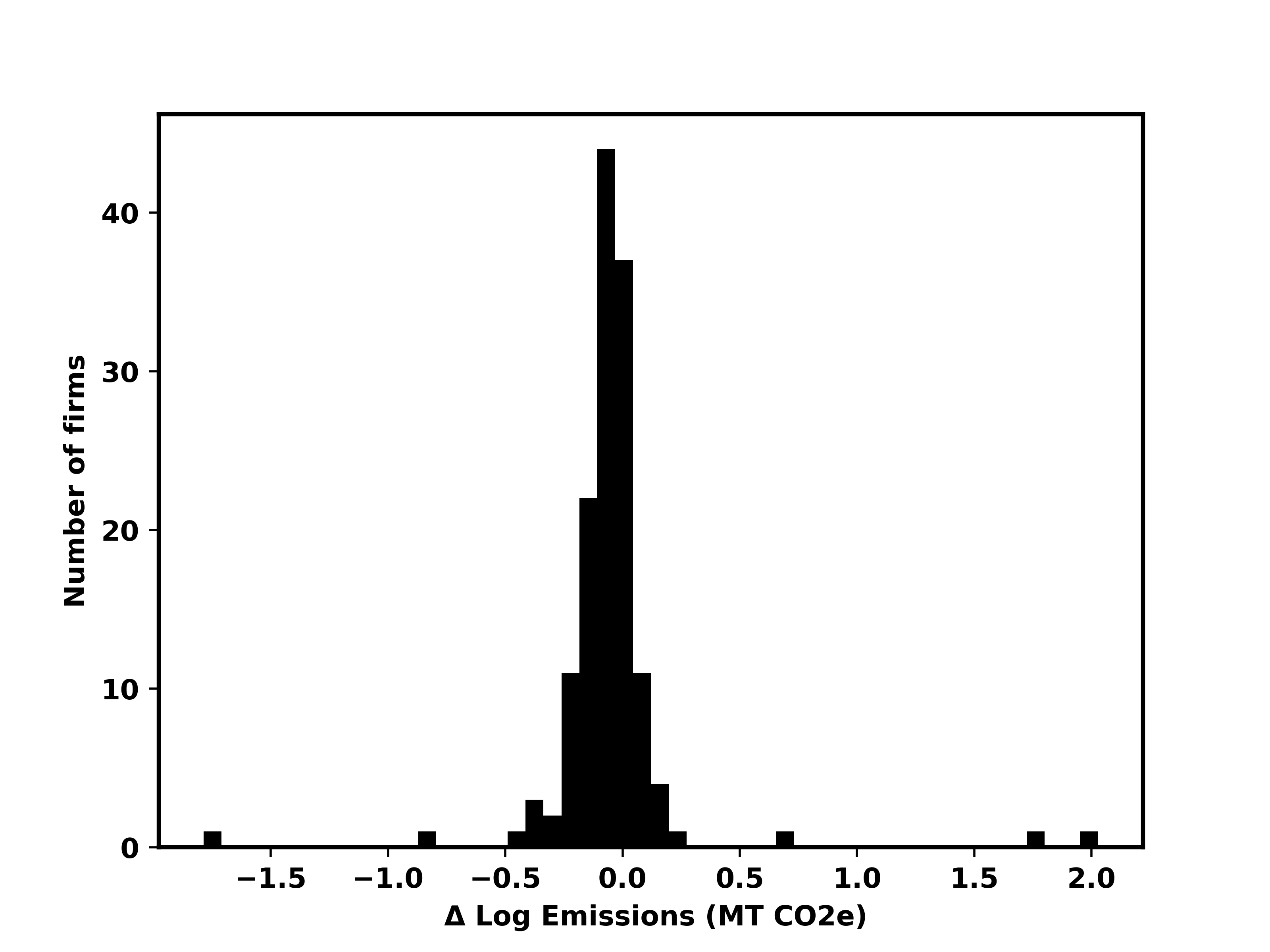}
\caption{Distribution of Changes in Log Emissions}
\label{subfig:em_hist}
\end{subfigure}
\begin{subfigure}{.5\textwidth}
\centering
\includegraphics[width=1\linewidth]{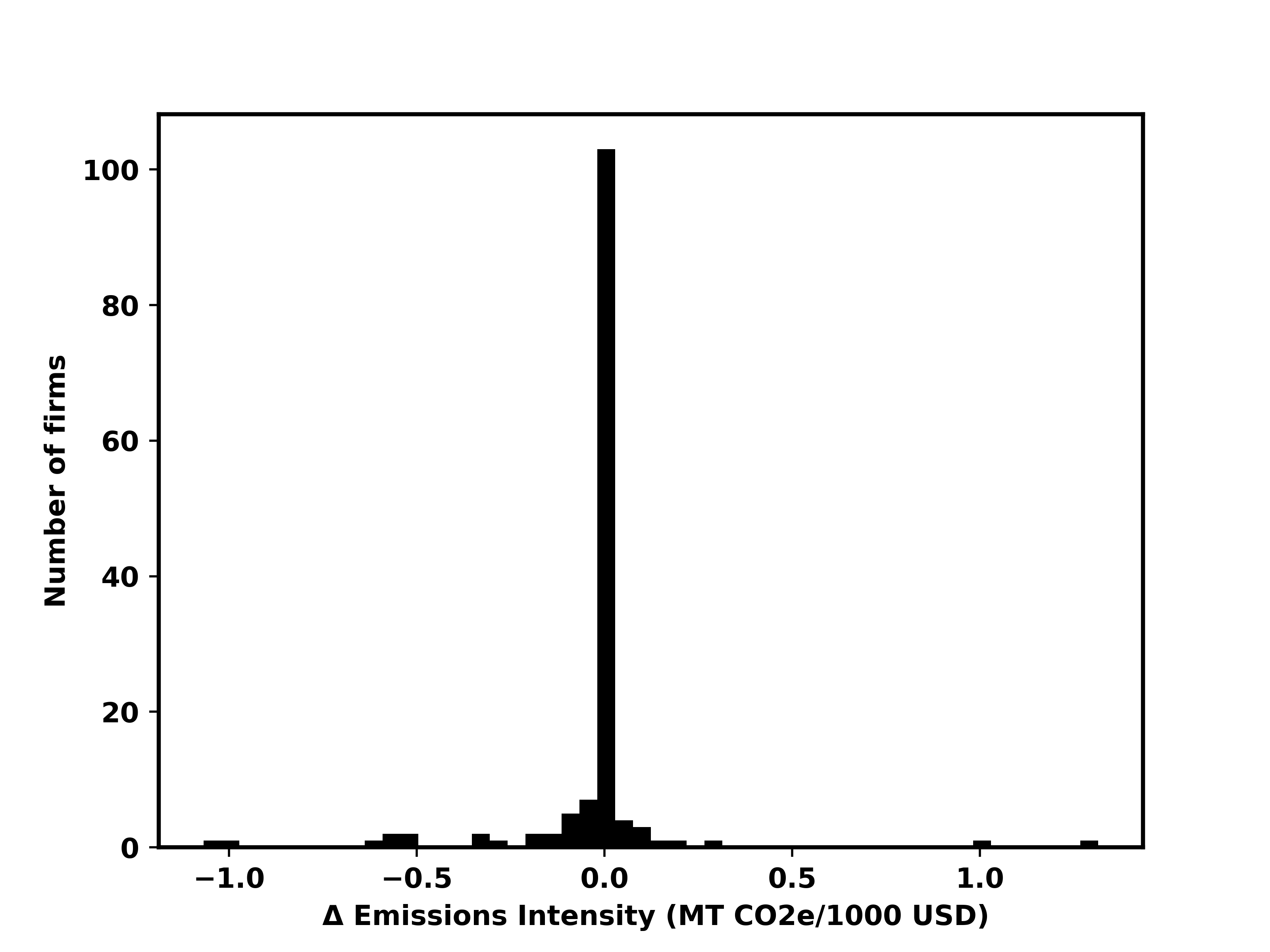}
\caption{Distribution of Changes in Emissions Intensity}
\label{subfig:int_hist}
\end{subfigure}
\caption{The distributions of changes in emissions (log) and changes in emissions intensity between 2018 and 2019. Emissions intensity is calculated by dividing GHG emissions by firm revenues of the same year. Includes 141 firms from 11 sectors.}
\label{fig:hist}
\end{figure}

\begin{figure}[h]
\begin{subfigure}{.5\textwidth}
\centering
\includegraphics[width=1\linewidth]{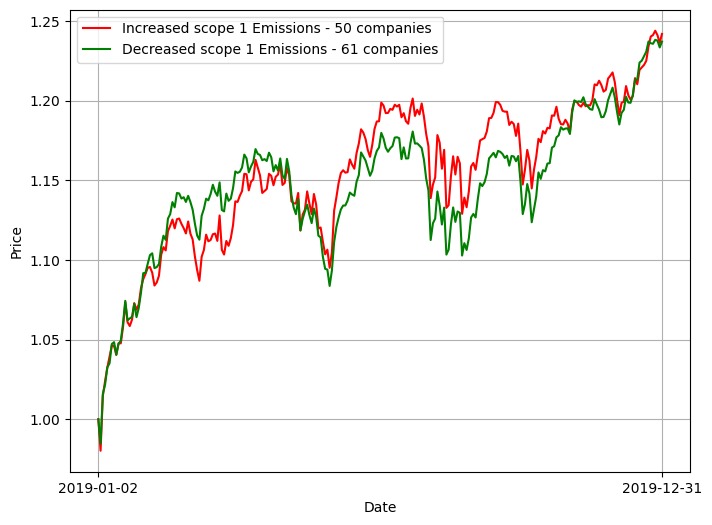}
\caption{Classified by Scope 1 emissions changes}
\label{subfig:s1changes}
\end{subfigure}
\begin{subfigure}{.5\textwidth}
\centering
\includegraphics[width=1\linewidth]{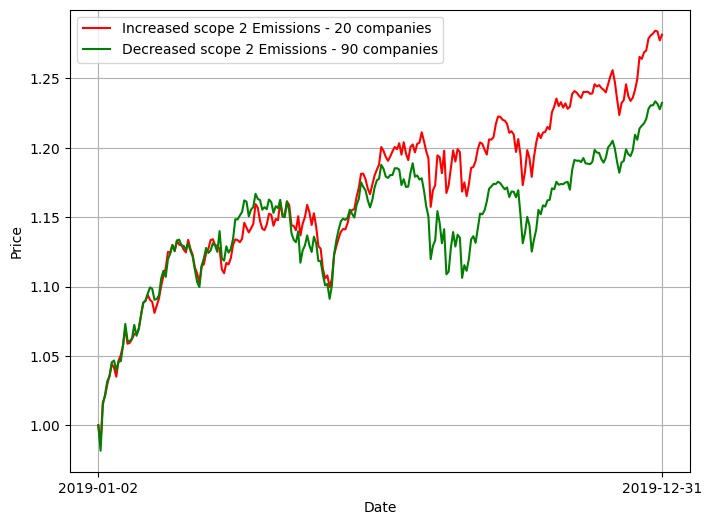}
\caption{Classified by Scope 2 emissions changes}
\label{subfig:s2changes}
\end{subfigure}
\caption{Stock prices of firms classified by (a) Scope 1 emissions changes, and (b) Scope 2 changes, for the year 2018-2019.}
\label{fig:trends}
\end{figure}

Table \ref{tab:sector_emissions} presents the Scope 1, Scope 2, and total (Scope 1 + Scope 2) emissions for firms from all sectors. Overall, emissions were reduced between 2018 and 2019, with an average reduction of 4.65\% in aggregate emissions. However, most of the reduction is from Scope 2 emissions, which declined by close to 18\%. On the contrary, Scope 1 emissions marginally increased by 1.08\% over 2018 levels. Within sectors, the biggest aggregate reduction in emissions is in the Consumer Staples sector with close to a 25\% reduction in emissions, while Materials and Health Care industries observe only marginal reductions of under 3\% each. While most sectors also reduced their Scope 1 emissions, the Information Technology sector increased their Scope 1 emissions by over 25\% above their 2018 level, probably due to increased production, which biases the average for all sectors combined. An implication of the large fall in Scope 2 emissions, against the small rise in Scope 1 emissions is that companies reduced their energy consumption substantially. The smaller reduction in Scope 1 emissions implies that the production of emissions from companies changes little in comparison (to bought emissions). This may be a consequence of businesses opting for more responsible electricity consumption as a means of reducing their carbon footprints, as opposed to significantly improving their production mechanisms. Figure \ref{fig:hist} displays the distribution of changes in aggregate emissions (subfigure \ref{subfig:em_hist}) and emissions intensities (subfigure \ref{subfig:int_hist}) of firms in the sample.\footnote{Emissions intensity is calculated by scaling aggregate annual emissions by the revenue generated in that year}. In both cases, we find tall peaks at the center, with few outliers, suggesting that most firms have comparable emission and electricity consumption patterns. However, the distribution of emissions intensity is considerably more skewed with a very tall peak around 0, suggesting that it may be more susceptible to being biased by outliers. The spread of changes in emissions is less skewed, and would make for a better treatment variable. Figure \ref{fig:trends} compares the stock prices (initialized at 1 USD at the start of 2018) between firms that increase and decrease their Scope 1 and Scope 2 emissions. From first glance, it appears as though prices are independent of whether firms increase or decrease their Scope 1 emissions, while firms that increase Scope 2 emissions fare better against firms that reduce Scope 2 emissions. However, there could be correlations with several other variables and a more robust empirical strategy which controls for other variables and firm-specific effects would be required for an adequate estimation of carbon premia. 
 
\begin{table}[h]
\centering
\resizebox{\textwidth}{!}{\begin{tabular}{ccccccccc}
\hline 
&&&&&&&&\\
& \multicolumn{8}{c}{Means (standard deviation)}  \\
\\ \cline{2-9}
&&&&&&&&\\
& \multicolumn{2}{c}{2018}&&\multicolumn{2}{c}{2019}&&\multicolumn{2}{c}{$\Delta$}\\
&&&&&&&&\\
\cline{2-3} \cline{5-6} \cline{8-9}
&&&&&&&&\\
& Emission Reducers &Emission Increasers && Emission Reducers &Emission Increasers && Emission Reducers &Emission Increasers \\
&&&&&&&&\\
\hline 
&&&&&&&&\\
Log Total Emissions (MT CO2e)&  13.7206     &  13.8380   &&  13.5977      &14.0185   &&  -0.1229 & 0.1806
\\
&     (2.4297)    &   (2.3299)    &&    (2.4228)     &    (2.3059)     &&   (0.1977)     & (0.4387)
\\
Total Liabilities (Billion USD)&  47.4739      &  33.5895     &&  50.7103       &34.6071     &&   3.2364   & 1.0176
\\
&     (91.6004)    &   (52.9324)    &&    (97.8752)     &    (52.4365)     &&   (8.9561)     & (4.6629)
\\
Total Assets (Billion USD)&  64.9555   &  49.2088   &&  68.8202   &  50.4200      &&  3.8646  & 1.2112
\\
&     (110.9986)    &   (75.0238)    &&    (117.7552)     &    (74.9736)     &&   (11.5056)     & (9.3826)
\\
Free Cash Flow (Billion USD)&  2.2361    &   3.7334   &&   2.7821      &   3.5361      &&  0.5460      & -0.1973
\\
&     (4.1155)    &   (8.8737)    &&    (7.1775)     &    (8.7526)     &&   (5.5011)     & (2.0240)
\\
Revenue (Billion USD)&    32.3890    &   36.8028    &&   32.7859     &  40.9709       && 0.3969      & 4.1680
\\
&     (47.6884)    &   (64.2901)    &&    (49.2913)     &    (68.7920)     &&   (7.7173)     & (17.8683)
\\
Operating Expenses (Billion USD)& 5.9696       &   5.9229    &&   6.3440     &    6.4711     &&    0.3744    & 0.5482
\\
&     (10.6697)    &   (9.3548)    &&    (11.2890)     &    (10.1253)     &&   (1.6191)     & (1.1386)
\\
EBITDA (Billion USD)&  5.8406       &   7.3910    &&  5.6491       &   7.7799      && -0.1915      & 0.3889
\\
&     (8.0379)    &   (15.0582)    &&    (7.3249)     &    (15.0398)     &&   (1.4937)     & (2.5904)
\\
Earnings per Share (USD)&  5.8920       &   5.4490    &&   6.1197     &  5.6683       &&  0.2277      & 0.3860
\\
&     (4.5913)    &   (4.3460)    &&    (5.2450)     &    (4.3801)     &&   (1.2170)     & (1.4849)
\\
Volatility &  0.0164   &   0.0174    &&   0.0150     &  0.0178       && -0.0015      & 0.0004
\\
&     (0.0044)    &   (0.0049)    &&    (0.0050)     &    (0.0062)     &&   (0.0029)     & (0.0038)
\\
&&&&&&&&\\
\hline 
\end{tabular}}
\caption{Financial characteristics of firms classified as Emission Reducers and Emission Increasers, in 2018 and 2019. Emission Reducers (106 firms) are defined as all those firms whose total (Scope 1 + Scope 2) emissions do not increase between 2018 and 2019, while the rest are defined as Emission Increases (35 firms). Includes 141 firms from 11 sectors. Averages are unweighted.}
\label{tab:desc_stats}
\end{table}

Table \ref{tab:desc_stats} presents the means and standard deviations of financial characteristics for the firms in the sample disaggregated by whether a firm's emissions increased between 2018 and 2019 (``Emission Increaser'') or whether its emissions decreased between 2018 and 2019 (``Emission Reducer''). The sample is split in almost exactly a three-to-one ratio in favor of emission reducers. This indicates that not only is the market getting cleaner overall, but the phenomenon is not due to a minority of firms cutting down on emissions by large amounts, but rather the dominant firm behavior. There are considerable disparities in the characteristics between Emission Reducers and Emission Increasers. As of 2018, Emission Reducers have more liabilities, less free cash flow, less revenue despite marginally higher operating expenses, and as a consequence lower EBITDA. However, these companies also have more assets,  higher earnings per share, and slightly lower volatility.\footnote{We define volatility to be the standard deviation of the daily log returns over one fiscal year.} Emission reducers have more assets and liabilities, but emission increases have higher revenue, profits, and cash flow. This suggests that firms that have high revenue and profits, may be less incentivized to cut down on their emissions due to high financial performance, although sectoral differences in Reducers and Increasers could also play an important role.\footnote{4 of 11 Energy companies increased emissions, while the numbers were 4(14) for Materials, 4(20) for Industrials, 3(9) for Consumer Discretionary, 2(7) for Consumer Staples, 5(20) for Health Care, 0(12) for Financials, 8(22) for Information Technology, 0(1) for Communication Services, 3(20) for Utilities, and 2(5) for Real Estate.}
If existing firm characteristics influenced whether a firm increased their emissions or not, then a selection bias in assignment to treatment must be corrected for to study whether there exists a carbon premium, or not. 

\section{Identification Strategy} 
\label{sec:method}

\subsection{Difference-in-differences design}

In order to study the effect of GHG emissions on stock prices, we begin with a two-way fixed effects model \citep{roth2023s}, with the logarithm of returns $r_{ijt}$ of firm $i$ of sector $j$ at time $t$, as the outcome variable: 
\begin{equation} \label{eq:TWFE}
r_{ijt} = \mu_{i} + \alpha_{t} + \theta_j + \beta \cdot  E_{ijt} + \gamma\cdot X_{ijt} + \epsilon_{ijt}, ~~~ t=2018, 2019
\end{equation}
Here, $\mu_i$ are firm-specific effects, $\alpha_t$ are time-specific effects, $\theta_j$ are sector-specific effects, $E_{ijt}$ is the vector of firms' Scope 1 and Scope 2 emissions, while $\beta$ is the vector of coefficients that measures the impact of the firms' Scope 1 and Scope 2 emissions, and $\gamma_{ijt}$ is the vector of coefficients of the control variables in vector $X_{ijt}$, and finally, $\epsilon_{ijt}$ is the idiosyncratic error term. Here, if $\beta$ is estimated as a statistically significant positive constant, it would imply that a positive carbon premium exists in the market. On the contrary, if $\beta$ is found to be negative, it means an equity greenium exists. Note that while we have data on $r_{ijt}, E_{ijt}$, and $X_{ijt}$, we do not have data on $\alpha_t, \mu_i,$ and $\theta_j$. Since $\mu_i$ and $\theta_j$, however, are time-invariant effects, we can eliminate these terms by taking the first difference of equation \ref{eq:TWFE}, and arrive at
\begin{equation} \label{eq:DID}
\Delta r_{ijt} = r_{ijt} - r_{ijt-1} = \Delta \alpha_{t} + \beta \cdot \Delta E_{ijt} + \gamma\cdot \Delta X_{ijt} + \Delta \epsilon_{ijt}, ~~~ t= 2019
\end{equation}
Since we have data for all of $\Delta r_{ijt}, \Delta E_{ijt},$ and $\Delta X_{ijt}$, we can estimate $\beta$ and $\gamma$ using ordinary least squares (OLS), with $\Delta \alpha_t$ as the intercept. 

\subsection{Inverse propensity score weighting}

A simple difference-in-differences specification assumes that the assignment of treatment is ``as good as random", i.e., whether a firm increases or reduces its emissions, and by how much, can be assumed to be determined randomly. However, this need not be true. Without additional tests, we cannot ascertain whether emissions cause an increase in stock prices or whether it is those firms that bring high returns to investors that are more lax about environmental governance. While the two are not mutually exclusive, it is imperative to correct for the latter while estimating the first, i.e., we need to ensure that there is no selection bias in the assignment of firms to treatment. 

One approach to control for selection bias in assignment to treatment is to weight observations by the inverse of their propensity scores to receiving the treatment that they received. The rationale behind controlling for selection bias assumes that firms are not ``treated" at random, i.e., whether a firm increases or decreases emissions is not random and may depend on the firm's existing covariates. 
Propensity scores are estimated to indicate the probability of a firm receiving the treatment that it got based on observable covariates, dividing by which results in a quasi-experimental set-up where selection bias is eliminated. The intuitive basis for inverse propensity score weighting is as follows: In a hypothetical case with a binary treatment, suppose, firms are classified into those that have profited in the preceding year, and those that have made losses. Suppose a profiting firm is twice as likely to increase emissions as a loss-making firm, i.e, say, the probability of inclusion to the treatment group for profit-making firms is 0.5, while that of loss-making firms is 0.25. Thus, there would be twice as many profit-making firms that increased emissions, when compared to loss making firms, which makes the treatment non-random. Thus, by weighting each firm by the inverse of the propensity score, here, assigning a weight of 2 for profit-making firms, and a weight of 4 for loss-making firms, both type of firms are equally likely to be represented in the treatment group. Similarly, in the control group, profit-making firms would have a higher weight, since they are normally less likely to be in the control group, in this example. 

In our case, the treatment variable is the change in the log emissions (see equation \ref{eq:DID}). Thus we have a case of estimating generalized propensity scores \citep{hirano2004propensity} with continuous treatment doses. Following \citet{hirano2004propensity}, we assume a normal distribution for treatment-given covariates. Let $Z_{ijt}$ be the vector of covariates that confound the assignment to treatment $\Delta E_{ijt}$, then the coefficients of confounders in determining the treatment can be estimated from the following equation: 
\begin{equation} \label{ps-ols}
\Delta E_{ijt} = \beta_0' + \beta_1' \cdot Z_{ijt} + \varepsilon_{ijt}
\end{equation}
where $\varepsilon$ has the distribution $\mathcal{N}(0,\sigma^2)$. Having estimated the coefficients of $\beta_0'$ and $\beta_1'$, we now have an estimate for the expected treatment dose $\hat{\Delta E_{ijt}}$ given by
\begin{equation}
\hat{\Delta E_{ijt}} = \beta_0' + \beta_1' \cdot Z_{ijt}
\end{equation}
Using ${\Delta E_{ijt}}$ and $\hat{\Delta E_{ijt}}$, we estimate the generalized propensity score $p_{i}$ for unit $i$ as
\begin{equation}
\hat{p_{i}} = \frac{1}{\sqrt{2\pi \hat{\sigma}^2}}\left(-\frac{1}{2\hat{\sigma}^2}\left(\Delta E_{ijt} - \hat{\Delta E_{ijt}} \right)^2 \right) 
\end{equation}
We can then weight each firm by 
\begin{equation}
w_i = \frac{1}{\hat{p_i}}
\end{equation}
Note that while the propensity-score-weighted-DID regression controls for selection bias due to observables, we still need to make the unconfoundedness assumption, i.e., we assume that there are no unobservable characteristics that induce bias in the treatment that each firm receives. 

\section{Results} 
\label{sec:results}

\subsection{What factors cause firms to increase or decrease emissions?}

For a simple difference-in-differences analysis, it is important that the treatment be random, i.e., changes in firms' carbon emissions would need to be exogenously determined by firm-level decisions alone independent of other covariates such as size, profitability, etc., which is unlikely to be true. Thus, we begin by studying how the changes in emissions of firms depend on firm characteristics in the pre-treatment period. Table \ref{tab:stage-1} presents the results of the effects of various 2018 levels of firm characteristics on subsequent changes in GHG emissions between 2018 and 2019 for both Scope 1 and Scope 2 emissions. We can observe that the change in Scope 2 emissions is not affected statistically significantly by any covariate. This implies that there may not be fixed reasons why firms change their Scope 2 emissions, and they may simply be determined by firm decisions or other unobservable characteristics, which we cannot test. 

\begin{table}[t]
\centering
\resizebox{0.8\textwidth}{!}{\begin{tabular}{cccc}
\hline 
\\
& \multicolumn{3}{c}{Coefficients (standard errors)}  \\
\\
\cline{2-4}
\\
&$\Delta$ Log Scope 1 Emissions  &&$\Delta$ Log Scope 2 Emissions \\ 
\\
\hline 
\\
Intercept     &  -1.4359*	  &&  -1.9809        \\
& (0.786)   &&    (2.825) \\
Log Total Liabilities     &  -0.2635**	  &&  -0.0762         \\
& (0.102)   &&    (0.367) \\
Log Total Assets       & 0.3344***	   && 0.0936      	    \\
& (0.116)   &&   (0.418)  \\
Free Cash Flow$^\dagger$     & -0.0871	   && 0.3957       \\
& (0.214)   &&   (0.769)  \\
Log Revenue       & -0.0088	   &&  0.1002         \\
& (0.062)   &&  (0.222)   \\
Operating Expenses$^\dagger$       & -0.1933	   &&  -0.5437        \\
& (0.226)   &&   (0.812)  \\
EBITDA$^\dagger$       & 0.3299	   &&  0.2967        \\
& (0.242)   &&   (0.870)  \\
Earnings per Share    $\times 10^{-7} $    &  -0.0015	  &&  0.0941   \\
& (0.084)   &&   (0.303)  \\
Volatility       &  12.4585*	  &&  0.6886      \\
& (6.607)   &&  (23.742)   \\
Log Scope 1 Emissions       &  -0.0467***	  &&  -0.0600       \\
& (0.014)   &&  (0.052)   \\
Log Scope 2 Emissions       & 0.0203	   &&   -0.0538       \\
& (0.017)   &&  (0.062)   \\
Momentum       & 0.0819	   &&  0.0227         \\
& (0.141)   &&  (0.508)   \\
Sector Fixed Effects & Yes && Yes\\
\\
\hline 
\end{tabular}}
\caption{The effect of 2018-levels of firm characteristics on the change in GHG emissions of firms between 2018 and 2019, over a sample of 141 firms indexed in S\&P500. Heteroskedasticity consistent standard errors. $^\dagger$Variable scaled by revenue.
\\ \small{***Significant at the 1\% level 
**Significant at the 5\% level
*Significant at the 10\% level}
}
\label{tab:stage-1}
\end{table}

On the other hand, changes in Scope 1 emissions are strongly correlated with several pre-treatment characteristics. A firm is likely to observe a more negative change in Scope 1 emissions if it has fewer assets, more liabilities, less volatility, and high existing Scope 1 emissions, with all effects being statistically significant (at least at the 5\% level). The last is of particular relevance because it implies that big polluters are self-correcting their emissions, although the effect itself is rather small. The EBITDA also has a positive effect on the change in Scope 1 emissions but is significant only at the 10\% level. However, since we study a relatively small sample of 141 firms, this effect cannot be dismissed entirely, as standard errors can be expected to be larger in smaller samples. Note that while Scope 2 emissions reduce by nearly 2\% on average (having controlled for observables), which is almost one and a half times the change for Scope 1 emissions, this effect is statistically insignificant. 

While these results are valuable in isolation as well, they are also important for the estimation of the effects of GHG emissions on stock returns. Since changes in Scope 1 emissions are not random, the causal effect cannot merely be studied through levels regressions, fixed effects regressions, or even simple difference-in-differences regressions, and the selection bias in assignment to treatment will need to be controlled for, before effects can be appropriately estimated. 
\subsection{Is there a carbon premium?}

Table \ref{tab:stage-2} presents the results of the unweighted (column 1) and weighted DID regressions, for the impact of emissions on stock returns, along with some important control variables, with column 2 presenting the results with Scope 1 emissions as the treatment and column 3 presenting the results with Scope 2 emissions as the treatment. The coefficients presented in column 1 and column 2, respectively, are our preferred estimate of the risk-premium associated with Scope 1 emissions and Scope 2 emissions. We find that there is a statistically significant (at the 1\% level) risk-premium associated with Scope 1 emissions. For every 1\% increase in Scope 1 emissions, there is a 0.076\% surplus return in the shares of a stock. The effect is of a similar size in the unweighted regression but is only significant at the 10\% level. This signifies that when firms that are less likely to increase emissions actually increase their emissions, there is a bigger rise in the stock price. This captures the effect of an unanticipated increase in a firm's polluting behavior, leading to a greater pricing of carbon risk in the market for said firm. 

\begin{table}[t]
\centering
\resizebox{1\textwidth}{!}{\begin{tabular}{cccc}
\hline 
\\
& \multicolumn{3}{c}{Coefficients (standard errors)}  \\
\\
\cline{2-4}
\\
& Unweighted DID  & \multicolumn{2}{c}{p-weighted DID}  \\ 
\\\cline{3-4}
\\
& &  Weighting: Scope 1 Emissions & Weighting: Scope 2 Emissions\\ 
\\
\cline{2-4} 
\\
& (1) & (2) & (3) \\
\\
\hline 
\\
Intercept     &  0.2176***     	  &   0.2191***         &    0.2453*           \\
& (0.0476) 	  &  (0.0466)      &  (0.1453)        \\
$\Delta$ Log Scope 1 Emissions       & 0.0629*      	  &   0.0761***         &   0.1113            \\
&  (0.0344)	  & (0.0245)      &    (0.1004)      \\
$\Delta$  Log Scope 2 Emissions       &  -0.0099     	  &  -0.0148          &  0.0114             \\
& (0.0356) 	  &  (0.0338)     &   (0.0110)       \\
$\Delta$  Log Total Liabilities       & -0.2230      	  &   -0.3530         &   -0.5456            \\
&  (0.2334)	  &  (0.2465)     &   (0.4463)       \\
$\Delta$  Log Total Assets         &  0.4574     	  &   0.6128*         &  0.9750             \\
&  (0.3144)	  &   (0.3191)    &  (0.6791)            \\
$\Delta$ Free Cash Flow &  -0.1709        & -0.1930        &  -0.1378       \\
&  (0.1207)   &  (0.1203)  &  (0.1657)  \\
$\Delta$  Log Revenue        &  -0.2775**     	  &  -0.2727**          &   -0.6478            \\
& (0.1175) 	  &   (0.1063)    &    (0.3953)      \\
$\Delta$ Operating Expenses &  0.0318        &  0.1701       &  -2.1917       \\
&  0.3961   & (0.3971)   &  (2.1872)  \\
$\Delta$ EBITDA &  0.3106        &  0.3432       & 1.2074        \\
&  (0.3658)   & (0.3771)   &  (1.0857)  \\
$\Delta$ Earnings per Share & 0.0000         &  0.0000       & 0.0000        \\
&  (0.0109)   &  (0.0068)  &  (0.0200)  \\
$\Delta$ Volatility &  1.0693        & 0.9510        &  -9.7905      \\
&  (5.6392)   & (5.1904)   &  (16.2962)  \\
Sector Fixed Effects & Yes & Yes & Yes\\
\\
\hline 
\end{tabular}}
\caption{The results for the impact of GHG emissions (carbon premium) and select control variables on stock returns of firms between 2018 and 2019, over a sample of 141 firms indexed in S\&P500. Heteroskedasticity consistent standard errors. 
\\ \small{***Significant at the 1\% level 
**Significant at the 5\% level
*Significant at the 10\% level}
}
\label{tab:stage-2}
\end{table}

However, we find that Scope 2 emissions do not produce a statistically significant risk premium, although the size of the premium is bigger. For every 1\% increase in Scope 2 emissions, there is a 0.111\% abnormal excess return on the stock that cannot be explained by other observable characteristics. However, due to high variation in the behavior of stocks, this effect is not significant. It is noteworthy that in the absence of weighting by propensity scores it may seem that there is a statistically insignificant equity greenium for Scope 2 stocks which would have led to a fall in stock prices for a rise in emissions, however, controlling for the assignment to treatment, the effect is found to be positive. 

\subsection{Is the carbon premium symmetric?}

It is possible that the carbon premium exists only for an increase in carbon emissions, and reductions aren't thought of to have significantly lead to a reduction in the price of the risk. To check this, we consider two treatment variables each for Scope 1 and Scope 2 emissions, allowing for separate carbon premia for emissions increasers and reducers.  
\begin{table}[t]
\centering
\resizebox{1\textwidth}{!}{\begin{tabular}{cccc}
\hline 
\\
& \multicolumn{3}{c}{Coefficients (standard errors)}  \\
\\
\cline{2-4}
\\
& Unweighted DID  & \multicolumn{2}{c}{p-weighted DID}  \\ 
\\\cline{3-4}
\\
& &  Weighting: Scope 1 Emissions & Weighting: Scope 2 Emissions\\ 
\\
\cline{2-4} 
\\
& (1) & (2) & (3) \\
\\
\hline 
\\
Intercept     &  0.2416***     	  &   0.2475***         &    0.3150***           \\
& (0.0522) 	  &  (0.0540)      &  (0.1084)        \\
$\Delta$ Log Scope 1 Emissions -- Increasers      & 0.0657**      	  &   0.0688***         &  0.0897             \\
&  (0.0301)	  & (0.0262)      &    (0.0717)      \\
$\Delta$ Log Scope 1 Emissions -- Decreasers     & 0.0175         	  &   0.1123            &   -0.1529               \\
&  (0.1426)	  & (0.1670)      &    (0.3258)      \\
$\Delta$  Log Scope 2 Emissions -- Increasers      &  -0.0991       	  &  -0.0909              &  -0.0806                 \\
& (0.1395) 	  &  (0.1557)     &   (0.2626)       \\
$\Delta$  Log Scope 2 Emissions -- Decreasers     &  0.0025     	  & 0.0007            &  0.0160              \\
& (0.0216) 	  &  (0.0244)     &   (0.0108)       \\
$\Delta$  Log Total Liabilities       &-0.1993      	  &   -0.2951            &   -0.2610                \\
&  (0.2349)	  &  (0.2436)     &   (0.3018)       \\
$\Delta$  Log Total Assets         &  0.4042     	  &   0.5069             & 0.4654                 \\
&  (0.3174)	  &   (0.3273)    &  (0.4049)            \\
$\Delta$ Free Cash Flow & -0.1785         &  -0.2023       &  -0.2140       \\
&  (0.1200)   &  (0.1262)  &  (0.1803)  \\
$\Delta$  Log Revenue        &  -0.2597**     	  &  -0.2406**          &   -0.5642              \\
& (0.1202) 	  &   (0.1100)    &    (0.3539)      \\
$\Delta$ Operating Expenses & -0.0018         & 0.2736        &  -1.5865       \\
&  (0.4414)   & (0.4356)    &   (2.0104) \\
$\Delta$ EBITDA & 0.2771         &  0.3999       &   1.2132      \\
&  (0.3863)   & (0.3910)   &   (1.1494) \\
$\Delta$ Earnings per Share  & 0.0000         & 0.0000        &  0.0000       \\
& (0.0074)    &  (0.0023)  &  (0.0113)  \\
$\Delta$ Volatility & 2.7817         &  2.1787       &  4.7551       \\
& (5.5513)    & (4.9504)   &   (7.6917) \\
Sector Fixed Effects & Yes & Yes & Yes\\
\\
\hline 
\end{tabular}}
\caption{The results for the impact of GHG emissions (carbon premium) and select control variables on stock returns of firms between 2018 and 2019, over a sample of 141 firms indexed in S\&P500. Heteroskedasticity consistent standard errors. 
\\ \small{***Significant at the 1\% level 
**Significant at the 5\% level
*Significant at the 10\% level}
}
\label{tab:stage-2-split}
\end{table}

Table \ref{tab:stage-2-split} presents the results for the estimation of carbon premia with different estimates for firms that increase and decrease their emissions. Column 2, our preferred estimate for the carbon premia associated with Scope 1 emissions, shows that carbon risk associated with Scope 1 emissions are only priced when emissions increase, and that there is no alleviation of carbon risk when emissions reduce at the rate at which they have, implying that emissions would have to reduce at a much faster rate in order to reduce risk for investors. Ideally, we would've expected to find a similar effect for Scope 2 emissions (Column 3), however, no effect on the premium is found on either increases or decreases in Scope 2 emissions. However, it is difficult to draw inferences from the estimates of Scope 2 emissions of emissions increasers as there are only 20 firms of this nature in the sample. 

\section{Conclusion} 
\label{sec:conclusion}

While several earlier studies have explored the presence of a carbon premium or equity greenium in financial markets, the causal aspect of the link has received little attention. In this study, we investigate whether there exists a causal link between emissions and the stock prices of firms. Using a panel of 141 stocks from the S\&P500 companies, we study if Scope 1 and Scope 2 emissions significantly impact stock prices. To deal with the problems of reverse causality and selection bias, we use a propensity-score-weighted difference-in-differences analysis to isolate the impact of emissions on prices. Our first stage regression finds that several firm characteristics such as a firms assets, liabilities, volatility and Scope 1 emissions have statistically significant effects on changes in firms' emissions. We then use these estimates to control for selection bias by propensity score weighting in the seconds stage. By scaling by propensity scores, we effectively balance the sample such that each firm is equally likely to receive a treatment (changes in emissions). Using a difference-in-differences regression which controls for sector, time, and firm-fixed effects, we find that there is a statistically significant, positive carbon premium associated with Scope 1 emissions. However, we do not find such an impact for Scope 2 emissions. We hypothesize that this may be because Scope 1 emissions increase on average, while Scope 2 emissions decrease, and an increase in emissions could lead to an increase in the price of carbon risk but a reduction at the current rate need not alleviate the risk. To test this, we estimate the coefficients separately for firms that increase emissions and those that decrease emissions. We find that for Scope 1 emissions which has a carbon premium overall, the carbon premium is largely due to firms that increase emissions and firms that decrease emissions do not have a carbon premium associated with them, which may confirm our hypothesis. 

Our results agree with the \citet{bolton2021investors} in the presence of a significant positive premium associated with Scope 1 emissions in US markets, but differ in the context of Scope 2 emissions. The difference may be due to a more rapid decarbonization of Scope 2 emissions in recent years reflected in the data used in our analysis compared to the flat or weak increases in Scope 2 emissions in the period considered by \citet{bolton2021investors}. It is also possible that our estimate of Scope 2 emissions is not significant due to a smaller sample of firms. Nevertheless, our results lend support to existing studies on carbon premia, by eliminating biases typically present in correlation regressions, by controlling for assignment to treatment. 

Our findings reveal key insights about the pricing of carbon risk in financial markets. First, we find that investors are aware of carbon risk and are demanding greater short-term returns from polluting firms. Secondly, we find that investors expect firms to reduce their emissions over time, and thus a reduction in emissions at the current rate does not alleviate risks associated with polluting firms. This implies that firms must decarbonize at a much faster rate to reduce investor risks. On the contrary, increases in firms's emissions leads to a substantial increase in the carbon risk which is demonstrated by the presence of carbon premia for stocks of companies that increase their emissions. 

\singlespacing
\setlength\bibsep{0pt}

\bibliography{bibliography.bib}

\end{document}